\documentclass{ws-ijmpa}
\usepackage[super,compress]{cite}
\usepackage{graphicx}
\usepackage{textcomp}
\begin{document}
\markboth{E. Torassa}{TOP detector for particle identification at Belle II}

%
\catchline{}{}{}{}{}
%

\title{TOP detector for particle identification at Belle II}

\author{Ezio Torassa \\ ( On behalf of the Belle II TOP group)}

\address{INFN Sezione di Padova\\
35131 Padova, Italy\\
ezio.torassa@pd.infn.it}

\maketitle

\begin{history}
\received{Day Month Year}
\revised{Day Month Year}
\end{history}

\begin{abstract}
The Time-Of-Propagation (TOP) counter is a ring-imaging Cherenkov detector designed to identify the charged hadrons in the barrel region of the Belle II detector.
The Belle II experiment collected data delivered by the SuperKEKB accelerator from March 2019 to June 2022.
After the first run period (Run1) a long shutdown (LS1) was dedicated to implement several accelerator and detector upgrades. The second collision period (Run2), started in January 2024. The components of the TOP detector and the performance during the Run1 will be reported,  the detector upgrade during LS1 and future upgrades will be described.    
\end{abstract}

\ccode{PACS numbers:}

\section{Introduction}	
\begin{figure}[hb]
\begin{minipage}{0.49\linewidth}
    \centering
    \includegraphics[width=0.75\linewidth]{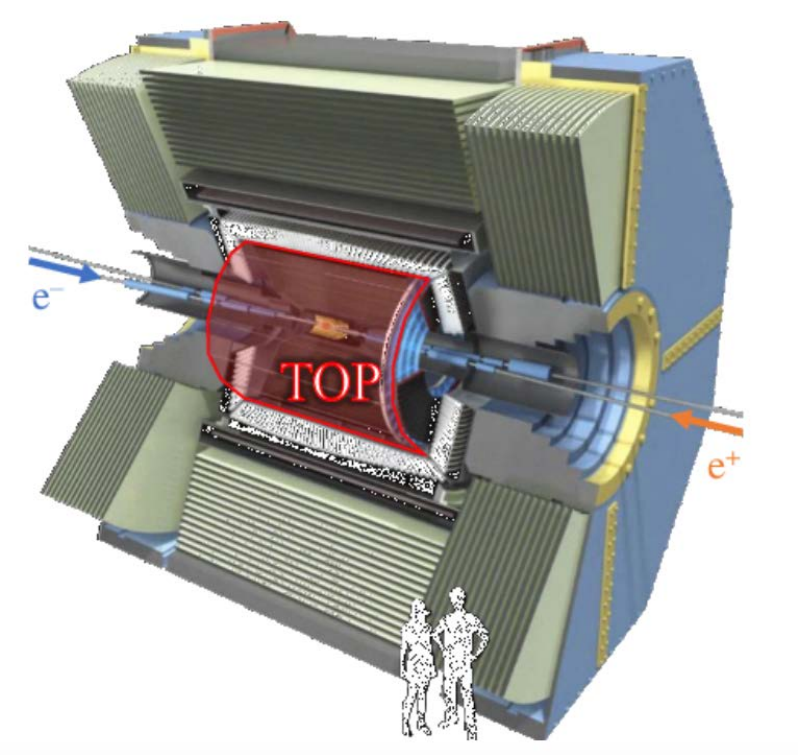}
\end{minipage}
\begin{minipage}{0.49\linewidth}
    \centering
    \includegraphics[width=0.90\linewidth]{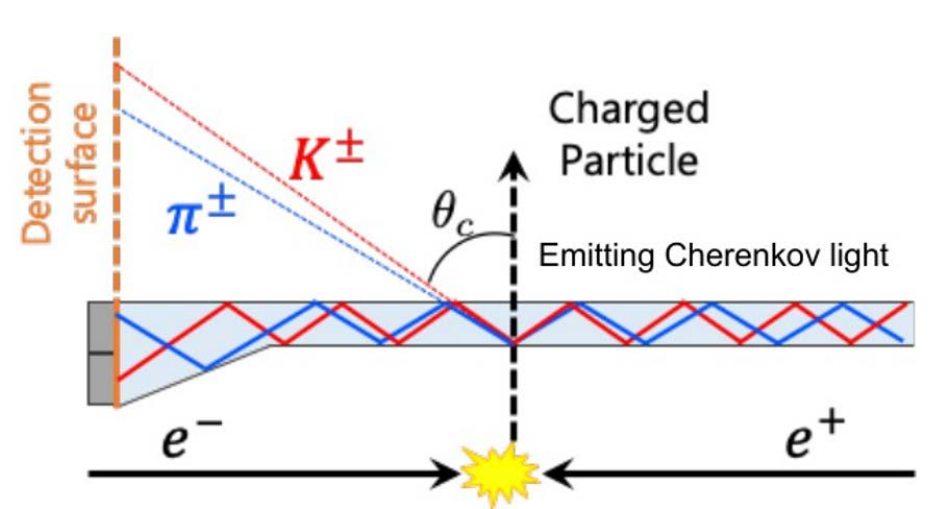}
\end{minipage}
\caption{The Belle II detector, whose size is 8 m wide and 8 m high, includes the TOP subdetector, whose dimensions are 2.7 m in length and 1.2 m in radius (left). Cherenkov photons emitted by pion and kaon hadrons at different angles and their arrival at the detection surface (right).}
\label{fig:f1}
\end{figure}

The TOP detector replaced the Aerogel Cherenkov Counter of the Belle experiment,
it consists of 16 modules made by quartz bar radiators read-out by microchannel plate photomultipliers  (MCP-PMTs).
The TOP modules are arranged in the central region of the Belle II detector \cite{b2det} between the Central Drift Chamber (CDC) and the Electromagnetic Calorimeter (ECL) (Fig.~\ref{fig:f1}, left) .
The quartz bars act as Cherenkov radiators and photon collectors. Thanks to the high average refractive index, the Cherenkov radiation remains trapped inside the bars and propagates to the photodetectors through internal reflection. Different hadrons crossing the quartz bar with the same angle have Cherenkov photons emitted at different angles, they arrive at the photodetector plane in different channels and at different times (Fig.~\ref{fig:f1}, right). 
The time of arrival is measured relative to the $e^+e^-$ collision time, it includes the time-of-flight (Tof) of the particle and the time-of-propagation (Top) of photons.
The Tof difference between $K$ and $\pi$ particles is related to the momentum $p$ and to the distance $L$ between the collision point and the entry point inside the TOP detector: 
$\Delta~\mathrm{Tof} =  L  (m^2_K-m^2_{\pi})/(2c\beta p^2)$. For $\beta\sim 1$, $p=2~\mathrm{GeV}$ and $L=1 ~\mathrm{m}$ the time-of-flight difference is $100~\mathrm{ps}$. 
The time-of-propagation difference between photons emitted by $K$ and $\pi$ particles is related to the distance they traveled  and to the difference between the Cherenkov angles, which in turn depend on the refractive index (n), the masses of the particles, and their momentum: $cos( \theta_c) = 1/\beta \mathrm{n} = \sqrt {(m^2+p^2)}/\mathrm{n}p$. The angle difference at $p= 2~\mathrm{GeV}$ is $1.55^o$, and $\Delta~\mathrm{Top} \sim 100~\mathrm {ps}$ \cite{Fast}.
The number of photons $N_{\gamma}$ is 20-30 depending on the track angle. 
The PID performance is proportional to $\sqrt{N_{\gamma}}/ \sigma_{Time}$ where $\sigma_{Time}$ is the total time resolution, including photodetector
and electronics time resolution. The arrival time and position of photons are measured with a 2D segmentation of 16 channels per $\mathrm{in}^2$ and a time resolution better than $100~\mathrm{ps}$. For every track, the probability of being a specific particle is estimated by comparing the resulting space-time distribution of the arriving photons with the probability density function expected for this particle hypothesis. The particle identification capability has been tested by selecting pure samples of pions, kaons, and protons tagged with reconstructing $D^*$, $K_S$ and $\Lambda$ particles  decays.

The TOP detector was constructed within the original time schedule; the installation was completed in May 2016. The drawback was the use of a small fraction of the latest generation of photomultipliers with a long lifetime. Half of the MCP-PMTs were replaced during LS1, and all others will be replaced during the next long shutdown. 

\section{TOP detector}
The key elements of the TOP detector are the Cherenkov radiator and light guide elements, the MCP-PMT photodetectors, the front-end readout electronics, and the laser calibration system.
\subsection {Cherenkov radiator and light guide}
\begin{figure*}[]
\centerline{\includegraphics[width=8.6cm]{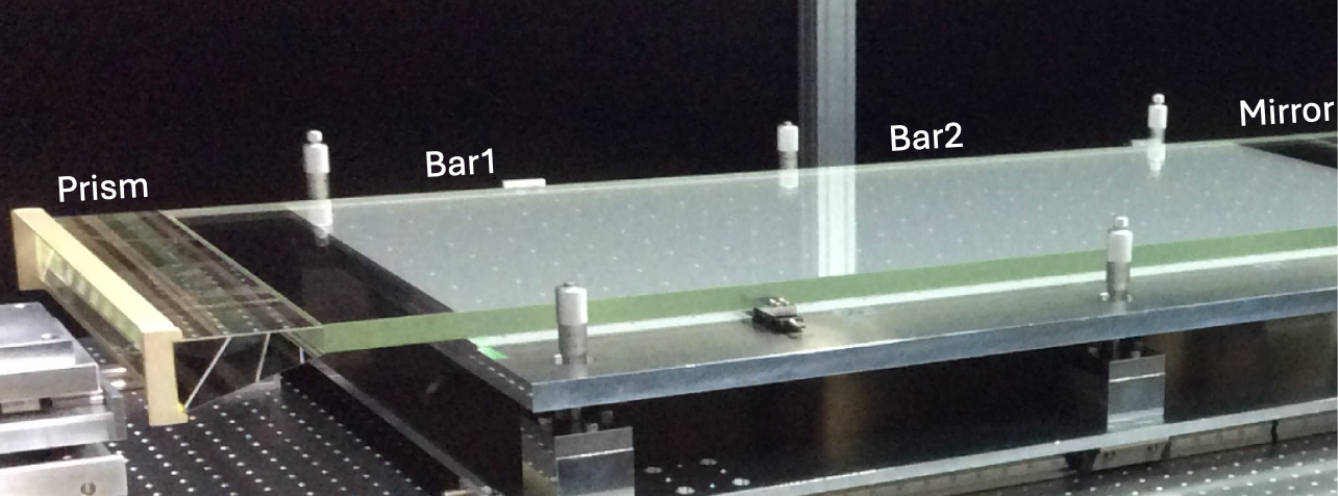}}
\caption{Cherenkov radiator and light guide elements for a single TOP module: two quartz bars glued together in the center, a quartz prism at the left, and a mirror at the right.} 
\label{fig:f2}
\end{figure*}

The Cherenkov radiator and light guide elements are shown in  Fig.~\ref{fig:f2}  for a single TOP module. 
Four parts are glued together: two fused silica bars each of dimension $(1250 \times 450 \times 20)~\mathrm{mm}^3$ acting as the Cherenkov radiator, 
a mirror located in the forward region with dimension $(100 \times  450 \times 20) ~\mathrm{mm}^3$ to reflect photons emitted forward, and a prism with dimension $(100 \times 456 \times 20-51) ~\mathrm{mm}^3$ acting as volume expansion, light guide, and couples the backward end of the quartz bar to an array of MCP-PMTs.
The high refractive index of the fused silica ($\mathrm{n}=1.47$ for $\lambda=400~\mathrm{nm}$) with high geometrical precision,
bulk transmittance, and surface reflectance (Table~\ref{ta1}) minimize the loss of photons during propagation \cite{quartz}. 

\begin{table*}[]
\tbl{Manufacturing specifications of the quartz bars.}
{\begin{tabular}{@{}ll@{}} \toprule
Quartz properties ~~~~~~~~~~~~~~~ & Requirements ~~~~~~~~~~~~~~~  \\ \toprule
Flatness & $< 6.3 ~\mu \mathrm{m}$  \\
Perpendicularity & $< 20 ~\mathrm{arcsec}$ \\
Parallelism & $< 4 ~\mathrm{arcsec}$ \\
Roughness & $< 0.5 ~\mathrm{nm}$ (RMS) \\
Bulk transmittance & $> 98\% / ~\mathrm{m}$ \\
Surface reflectance & $> 99.9\%$ / reflection \\ \botrule
\end{tabular} 
\label{ta1}}
\end{table*}

\subsection {MCP-PMT photodetectors}
\begin{figure}[h]
\begin{minipage}{.5\textwidth}
  \centering
  \includegraphics[width=.5\linewidth]{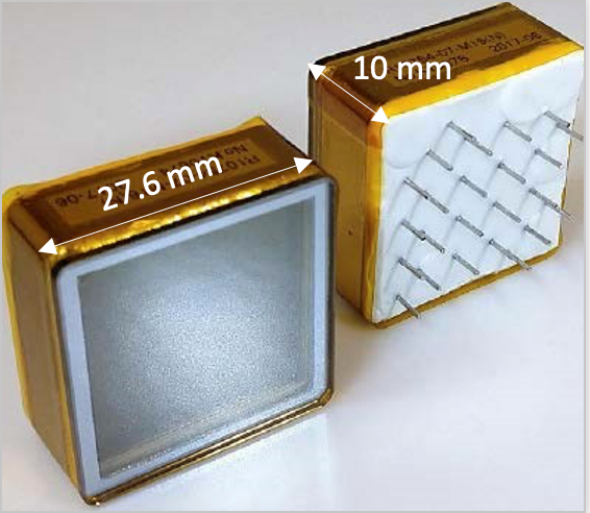}
\end{minipage}%
\begin{minipage}{.5\textwidth}
  \centering
  \includegraphics[width=0.6\linewidth]{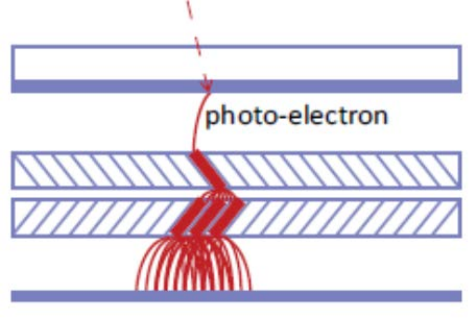}
\end{minipage}
\caption{The Hamamatsu R10754-07-M16(N) microchannel plate photomultiplier tube (left). Two microchannel plates with angled channels (right).}
  \label{fig:f3}
\end{figure}

The expansion volume of the quartz radiator bar is coupled on the back side with an array of two rows of MCP-PMTs, for a total of 32  photodetectors per module.  
Each MCP-PMT is square-shaped, has an effective area of $(23 \times 23)~ \mathrm{mm}^2$, and is segmented into 16 channels (Fig.~\ref{fig:f3},~left). The 16 TOP modules have a total of 8192 channels. The photoelectrons are multiplied inside two microchannel planes; each plane is $400~\mu \mathrm{m}$ thick, and each channel has a $10 ~\mu \mathrm{m}$  diameter. Channels are tilted to prevent electrons from passing through the holes without the multiplication process (Fig.~\ref{fig:f3},~right). 
The multiplication in the short path makes the transit time spread less than $50~\mathrm{ps}$.
The major problem with using MCP-PMTs is the quantum efficiency deterioration of the photocathode. Positive ions desorbed from the microchannel plate in the electron multiplication are scattered back to the photocathode, affecting the MCP-PMT lifetime. 
During mass production, three types of MCP-PMTs have been developed to improve the lifetime of the photocathode \cite{mcplife,mcpprod}. 
\begin{enumerate}
\item  Conventional MCP-PMT \\
In the manufacturing process of conventional MCP-PMT, a lead glass disk with holes is built with thermal and etching processes \cite{cmcp}.  
This lead glass capillary array is then heated in the presence of hydrogen, which chemically reduces the surfaces of holes, leaving a resistive and
emissive surface that is effective for electron amplification.
\item Atomic layer deposition ALD MCP-PMT \\
This new technology does not use materials containing lead, which is a harmful element restricted by the RoHS (restriction of hazardous substances) 
directive of the European Union. The new glass capillary array is coated using ALD technology with a resistive film and a secondary electron
multiplier film with high uniformity in each microchannel. 
\item Life-extended ALD MCP-PMT \\
Each component of the MCP-PMT gets an optimized bake-out to reduce the trapped ions. The residual ions can be released during the development 
of the electron avalanche and can reduce the lifetime of the photocathode.  
\end{enumerate}
Based on the development of the MCP-PMT technology during the TOP detector construction, $44\%$ of the installed PMTs were conventional MCP-PMTs, $43\%$ were ALD MCP-PMTs, and $13\%$ were life-extended ALD MCP-PMTs.
\subsection {Front-end readout electronics}
\begin{figure}[hbt]
\begin{minipage}{.5\textwidth}
  \centering
  \includegraphics[width=.55\linewidth]{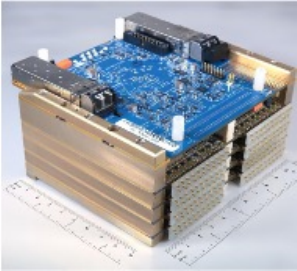}
\end{minipage}%
\begin{minipage}{.5\textwidth}
  \centering
  \includegraphics[width=0.85\linewidth]{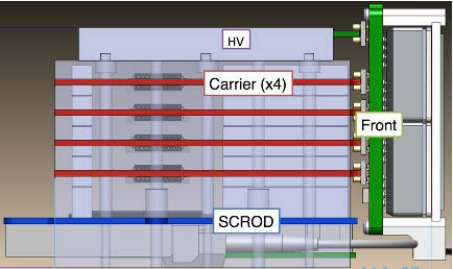}
\end{minipage}
  \caption{Assembled TOP boardstack (left). Drawing of the TOP boardstack components consisting of one SCROD board, four AISC boards, and one HV supply module (right).}{}
\label{fig:f4}
\end{figure}

The TOP front-end electronics are required to read out the signals of all 8192 MCP-PMT channels in the whole TOP system with a global timing resolution of better than $100 ~\mathrm{ps}$ at a nominal trigger rate of up to $30~ \mathrm{kHz}$.  This is achieved by employing specially designed $2.7~\mathrm{GHz}$ wave-form sampling electronics \cite{sampl}.
The TOP readout is organized as an ensemble of 4 boardstacks per module (Fig.~\ref{fig:f4}). Every boardstack contains 4 ASIC carrier boards and a single
Standard Control Read-Out data (SCROD). Every carrier board contains 4 Ice Ray Sampler version~X (IRSX) chips with 8 channels/chip \cite{feel}.
For every channel, an internal trigger system based on a fixed threshold marks regions of interest.
Every region of interest has a length of 32 samples, or about 12 ns.
The SCROD extracts the timing of photon pulses for every channel and transfers the data to the Belle II DAQ system.    

\subsection{Laser calibration system}
The laser calibration system was designed to calibrate the time differences between the 8192 channels of the TOP detector with a resolution better than $100~\mathrm{ps}$ \cite{calib}. To achieve this, it is necessary to reach all channels of all photodetectors with a single photon source while maintaining a time resolution of a few tens of picoseconds
in each step. To inject the calibration light into the box containing the quartz prism, the number and size of holes should be as minimal as possible to avoid the loss of Cherenkov light.
Simulation studies have shown the best optical and mechanical solution was to inject the light in 9 holes with steps of $50.7~\mathrm{mm}$, a vertical angle of $15^o$, and a numerical aperture $\mathrm {NA} \sim 0.5$--$0.6$, where $\mathrm{NA} = sin (\theta)$ and $\theta$ is the half angle of the light cone.
     
\begin{figure}[hb]
\begin{minipage}{.5\textwidth}
  \centering
  \includegraphics[width=1.2\linewidth]{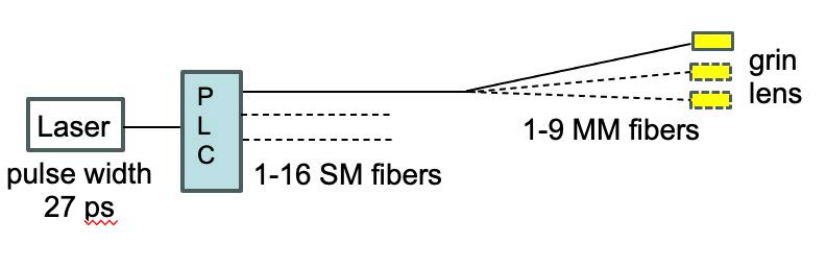}
\end{minipage}%
\begin{minipage}{.5\textwidth}
  \hspace*{1 cm}
  \centering
  \includegraphics[width=0.66\linewidth]{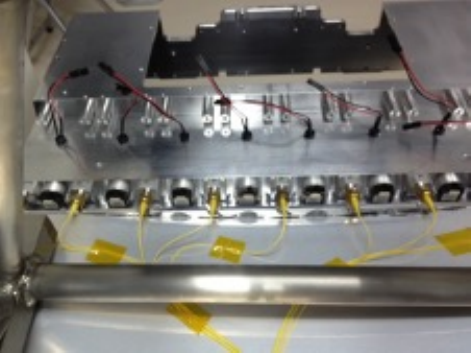}
\end{minipage}
  \caption{Scheme of the laser calibration system (left). Picture of the prism box with the fibers and grin lens of the laser calibration system installed.}{}
  \label{fig:f5}
\end{figure}

The calibration light source is provided by  a PiL040X pulsed laser diode with a wavelength of $405~\mathrm{nm}$, a pulse width less than $27~\mathrm{ps}$, and a peak power of $300~\mathrm{mW}$.
A Planar Lightwave Circuit (PLC High Power Splitter from PPI Inc.), followed by 30-meter-long single-mode fibers, demultiplexes the light for the 16 TOP modules.
For every module, a multi-mode fiber bundle splits the light into the nine light entries. For every bundle, the end side of every fiber is coupled with a grin lens (Selfoc-SLH) with a nominal $\mathrm{NA}=0.6$ (Fig.~\ref{fig:f5},~left).      

\section{Time calibration}
The aim of the time calibration is to even out the responses of the 8192 channels and, for every channel, equalize the responses of different sampling elements.  
Time differences may result from different cable routings and different electronic components. The time calibration is performed in four steps \cite{tcalib}:
\begin{enumerate}
\item {Time Base Calibration} \\
Each channel of the IRSX ASIC consists of 128 sample and hold cells, a circuit that has a 14 fF capacitor and a comparator, allowing the capture and storage of the time-sampled analog waveform. 
The sampling starts at the rising edge of the synchronization clock (accelerator clock divided by 24) and is performed at a rate of 2.7 GSa/s. 
The samples are transferred to 32k storage units. Different cells can have different delays to be precisely calibrated. 
A double-peak pulse with a fixed time difference of $22~\mathrm{ns}$ is injected into the electronics.
To cover all the 128 cells, the double pulse is not synchronized with the accelerator clock.
Figure~\ref{fig:f6} shows the measurement of the time difference as a function of the sample number. 
Time fluctuations up to $1~\mathrm{ns}$ have been reduced after the time base calibration to $42~\mathrm{ps}$ of r.m.s.
\begin{figure*}[h]
\begin{minipage}{.5\textwidth}
  \centering
  \includegraphics[width=0.94\linewidth]{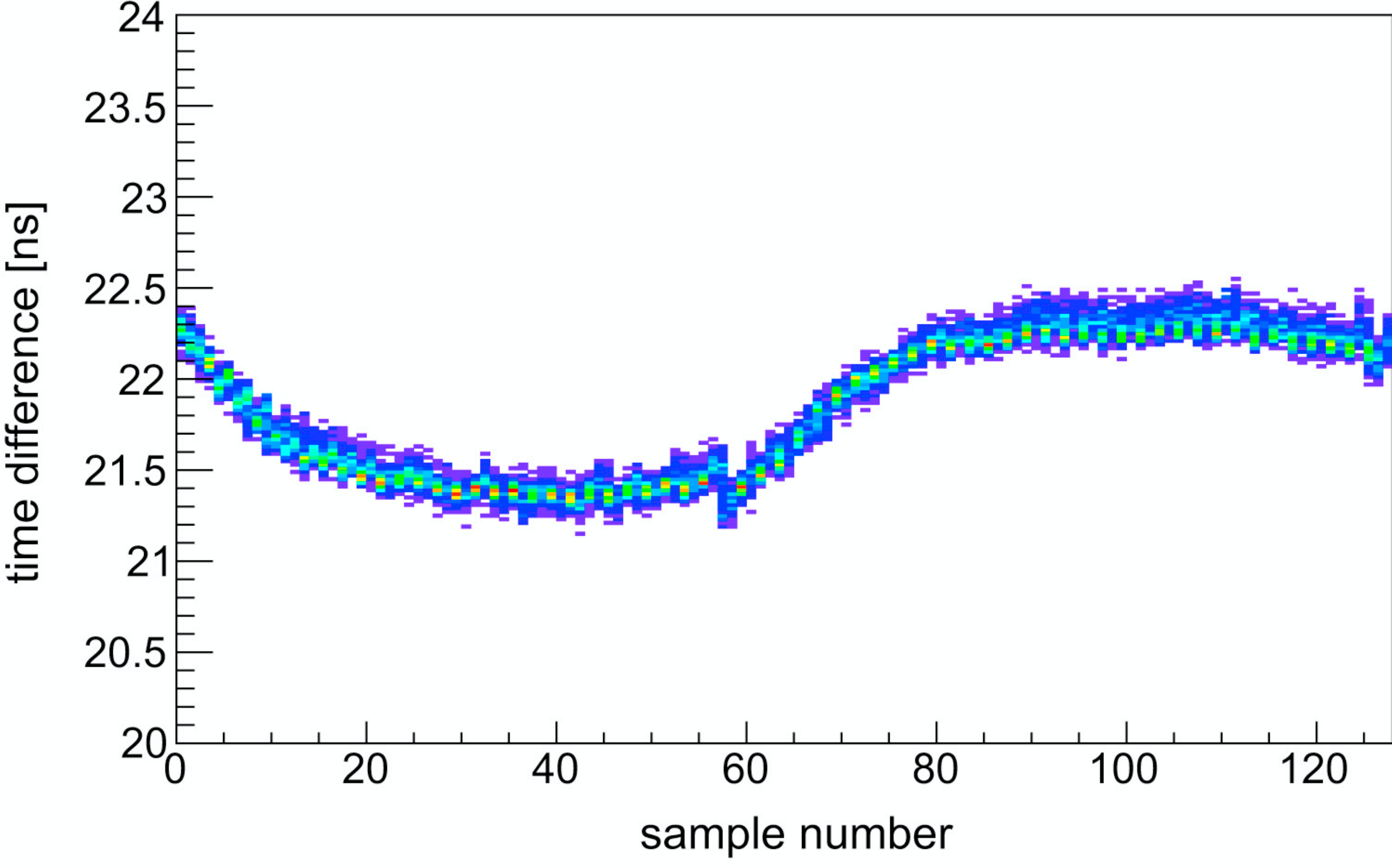}
\end{minipage}%
\begin{minipage}{.5\textwidth}
  \centering
  \includegraphics[width=0.94\linewidth]{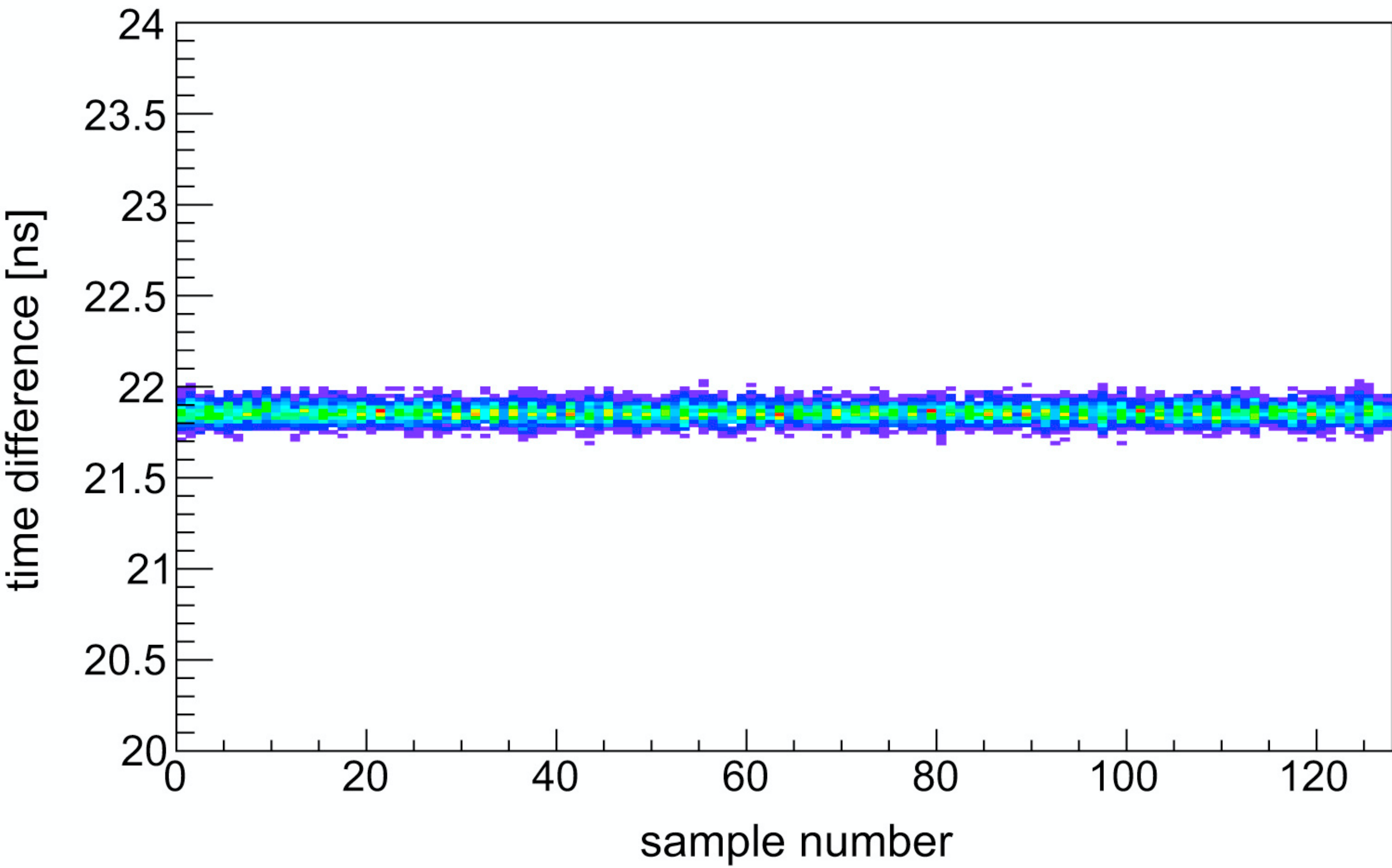}
\end{minipage}
\caption{The time difference between two calibration pulses as a function of sample number before (left) and after (right) the time base calibration.}
 \label{fig:f6}
\end{figure*}

\item {Time assignment of channels within a module} \\
The time alignment of channels within a module is performed with the laser calibration system.
The laser light is injected through the holes of the prims and reaches all 512 channels of the module at the same time.
The low timing jitter of the laser allows the different time responses of the channels to be measured.
However, single channels can receive light directly or after reflections inside the prism; they can also be illuminated by two fibers.
 A simple method is to align the average of the time distribution in each channel.
Figure~\ref{fig:f7} shows the times for the 128 channels in a single boardstack with respect to a common time related to the laser trigger before and after the alignment.
A more precise method with time offset fit was implemented; it considers the different intensities after reflections and the possible propagation times taken from Monte Carlo.
\begin{figure}[h]
\begin{minipage}{.5\textwidth}
  \centering
  \includegraphics[width=0.94\linewidth]{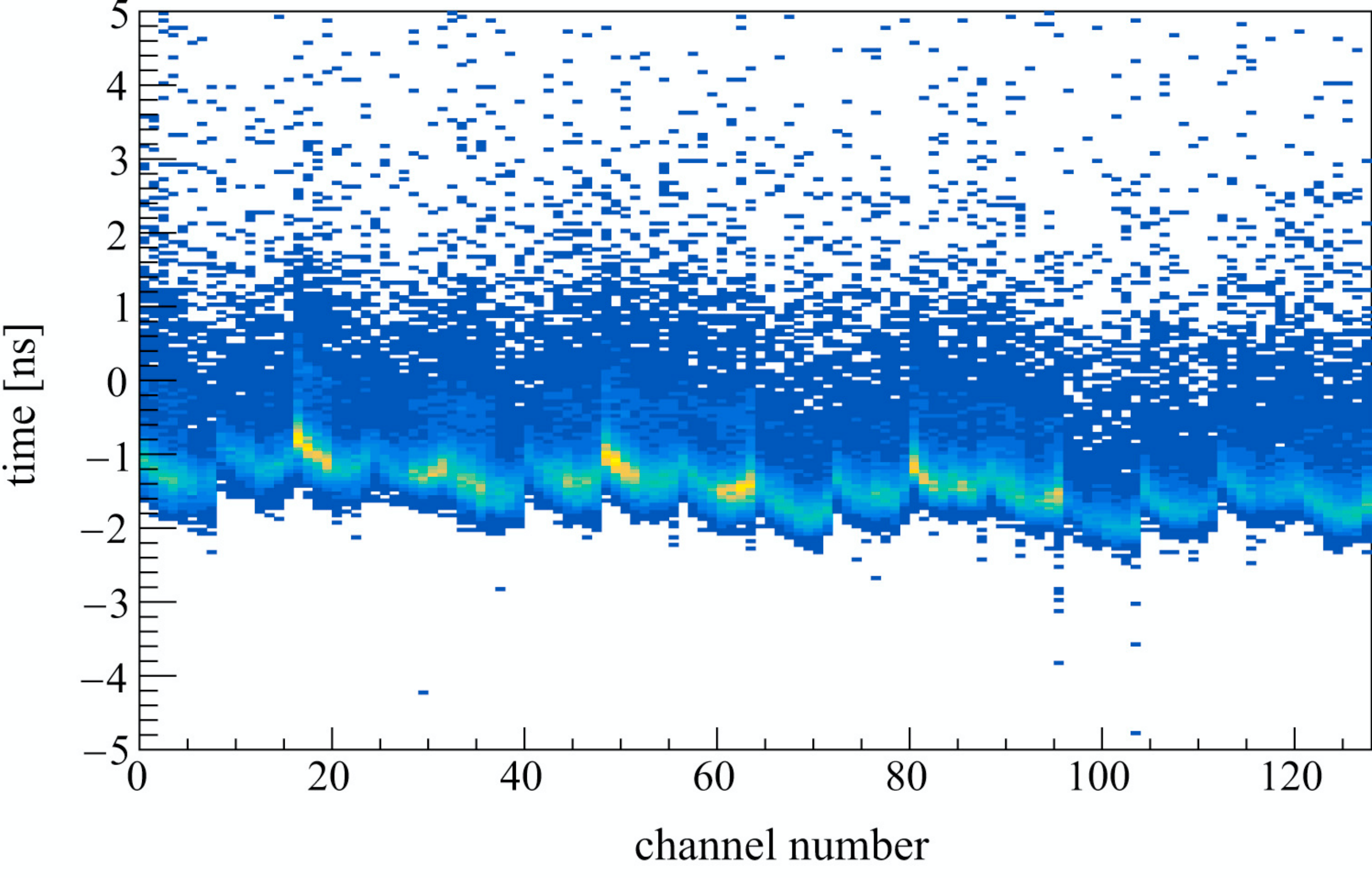}
\end{minipage}%
\begin{minipage}{.5\textwidth}
  \centering
  \includegraphics[width=0.94\linewidth]{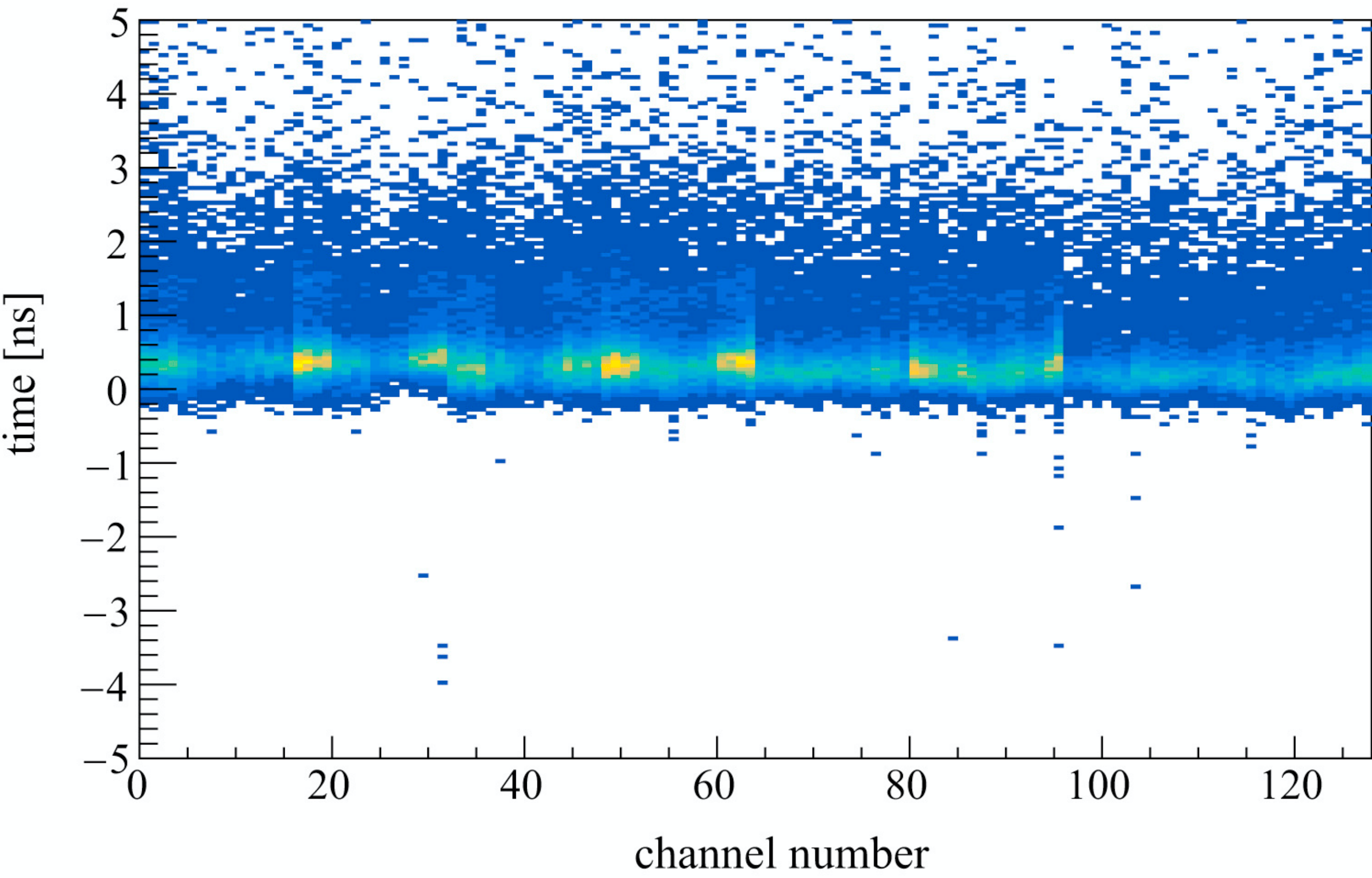}
\end{minipage}
\caption{Times for the 128 channels in a single boardstack with respect to a common time related to the laser trigger before (left) and after (right) alignment.}
 \label{fig:f7}
\end{figure}
\item {Time alignment of modules}  \\
The 16 TOP modules are aligned together within a few tens of ps using cosmic rays and dimuons from collision data.
The same kind of data can also be used for the geometrical alignment after summer and winter stops or after an earthquake.

\item {Alignment relative to collision time} \\
The global $T_0$ with respect to the RF accelerator clock is measured using dimuons from collision data.

\end{enumerate}

\section{TOP detector performance}
\begin{figure}[ht]
\centerline{\includegraphics[width=7.0cm]{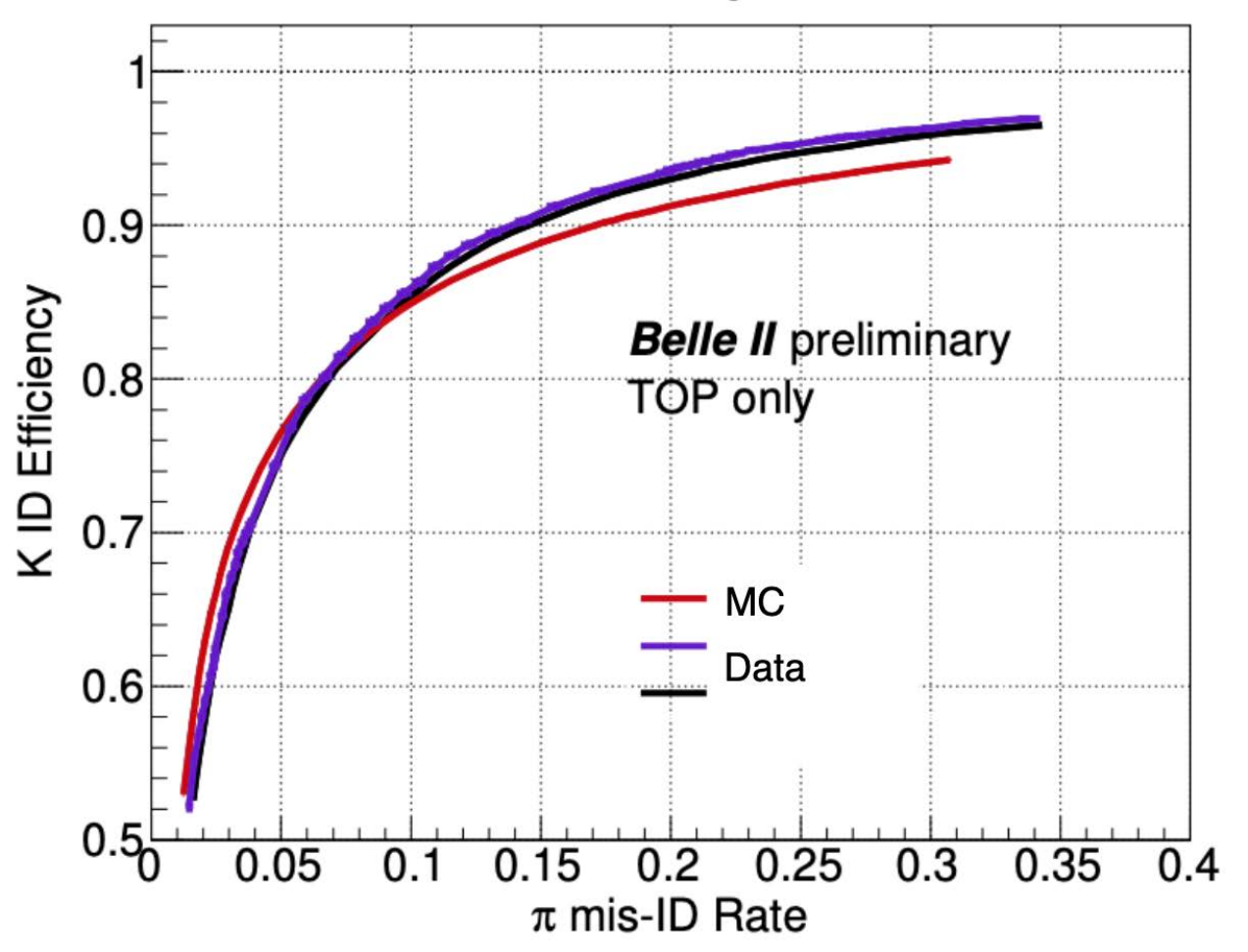}}
\caption{K identification efficiency as a function of the pion mis-ID rate for two data-set periods (blue and grey) and for the Monte Carlo (red). Lines join the selected points; the statistical error is on the order of \textperthousand .}
\label{fig:f8}
\end{figure}

\begin{figure*}[]
\begin{minipage}{.5\textwidth}
  \centering
  \includegraphics[width=0.95\linewidth]{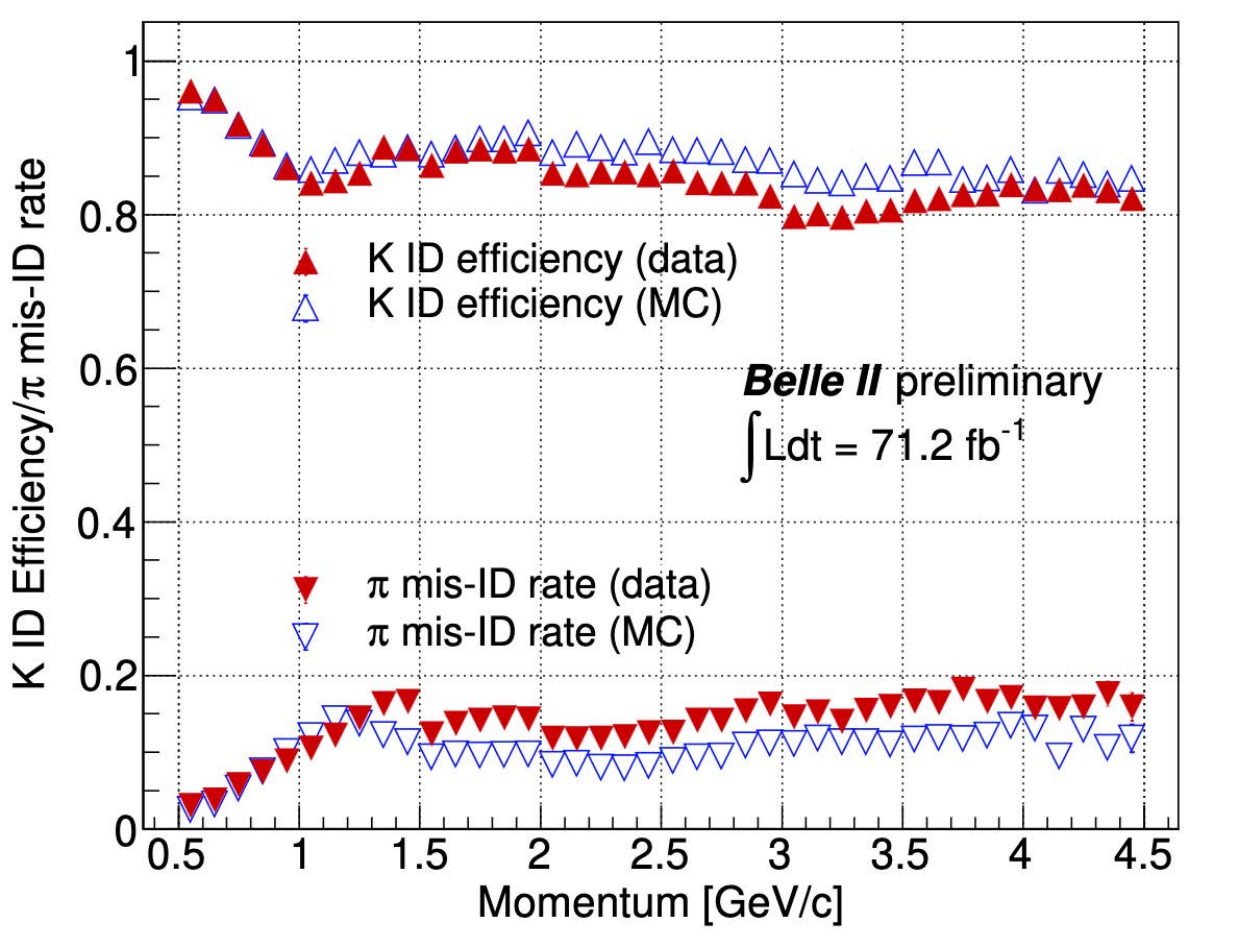}
\end{minipage}%
\begin{minipage}{.5\textwidth}
  \centering
  \includegraphics[width=0.95\linewidth]{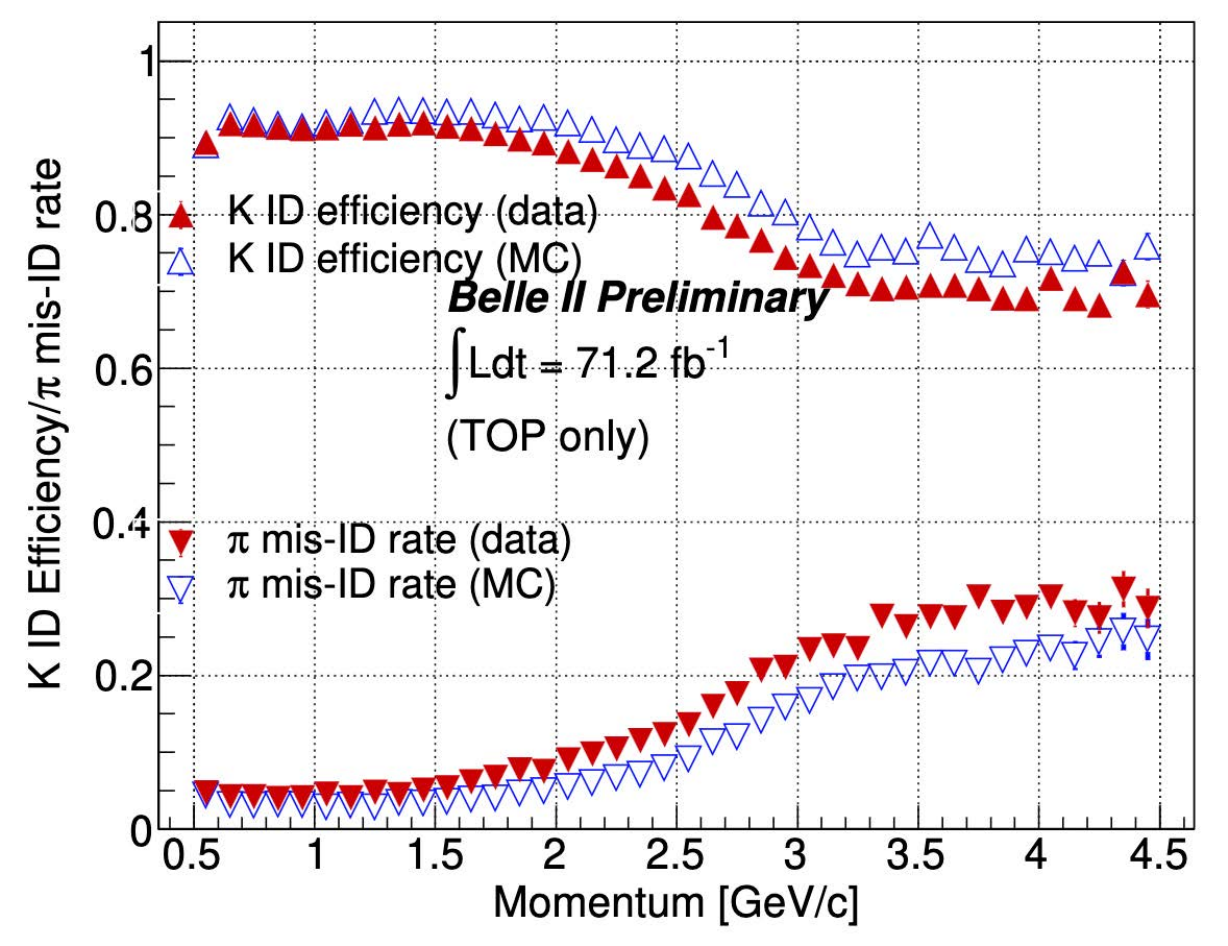}
\end{minipage}
\caption{Kaon identification efficiency and $\pi$ mis-ID rate as a function of momentum in the laboratory frame for the PID selection criterion $R[K/\pi ] > 0.5$ using all sub-detectors (left) or TOP only (right).}
  \label{fig:f9}
\end{figure*}

A relativistic charged track passing through the quartz bar produces N Cherenkov photons,  the $i$-photon arrives at the $c_i$ channel at the $t_i$ time.
The TOP likelihood is computed under the $\alpha$ particle hypothesis, like the following:
$${\mathcal{L}^{\mathrm{TOP}}_{\alpha}}=exp \Big[\sum^N_{i=1}log \Big (\frac{N_{\alpha}\cdot PDF^i_{\alpha}+N_{B}\cdot PDF^i_{B}}{N_{\alpha}+N_{B}}\Big ) +log P_N (N_{\alpha}+N_{B})\Big]$$
where $N_{\alpha}$ is the number of expected signal photons, $N_B$ is the number of expected background photons, $PDF^i$ is the probability density function for the $c_i$ channel at the $t_i$ time, and $P_N$ is the Poisson probability to detect N photons expecting $N_{\alpha}+N_{B}$.
Particle identification performance is estimated with a likelihood ratio by comparing the assumed particle with respect to the sum of alternative options. For a binary comparison between $K$ and $\pi$:
$$R[K/\pi]= \frac{\mathcal{L}_K}{{\mathcal{L}_K}{+\mathcal{L}_{\pi}}}$$
The likelihood ratio can be estimated not only for the TOP detector but also for all the other detectors providing information related to PID.
The PID performance of the TOP detector shown in Figure~\ref{fig:f8} was obtained by varying the $K-\pi$ binary ratio.
Two  data-set periods are reported and compared with the Monte Carlo expectation. 
The performance is stable over time; the difference with respect to Monte Carlo shows the simulation of the detector can still be improved. 
The identification efficiency for Kaons selected with $R(K/\pi ) > 0.5$ and the misidentification (mis-ID) rate of pions are shown in Figure~\ref{fig:f9} in bins of momentum using all
sub-detectors (left) or TOP only (right)  information \cite{perf}. Kaons and pions have been tagged from the $D^{*+} \to  D^0[K^-\pi^+]\pi^+$ decay.
The efficiency is about $85\%$ and almost uniform in all the TOP slots.
A machine learning approach is under study, where weights used to combine PID
information are not static but a function of the charge of the track and its kinematics.

\section{TOP detector upgrade}

After the first run period, 4 boardstacks were dead, and10 boardstacks were showing small issues during operation.
The TOP detector had, in total, about 600-700 dead channels out of 8192. All the boardstacks and electronics boards showing problems have been replaced during LS1.
The accumulated charge of the photodetectors was between $0.2~\mathrm{C/cm}^2$ and $0.3~\mathrm{C/cm}^2$, depending on the slot and the type of photodetector.
The lifetime of MCP-PMTs is defined as the accumulated charge, which has the effect of reducing the photocathode quantum efficiency (QE) to $80\%$.
The lifetimes measured in the laboratory were on average $1.1~\mathrm{C/cm}^2$ ($0.3$--$1.7 ~\mathrm{C/cm}^2$) for conventional MCP-PMTs,
$10.4~\mathrm{C/cm}^2$ ($2.5$--$26.1 ~\mathrm{C/cm}^2$) for ALD MCP-PMTs, and higher than $13.6~\mathrm{C/cm}^2$ ($13.6$--$33.4 ~\mathrm{C/cm}^2$)   for life-extended ALD MCP-PMTs \cite{mcplife}.
Considering the large fluctuation of these measurements and the expected increase in luminosity and background,
the conventional MCP-PMTs have been replaced, even though the QE was not expected to be significantly degraded on average.
The removed MCP-PMTs have been tested in the laboratory, and the average measured $\mathrm{QE_{rel.}}= \mathrm{QE_{Run1\textunderscore end}}/ \mathrm{QE_{Run1\textunderscore start}}$ was found to be 0.98 (expected 1) for life-extended ALD MCP-PMTs, 0.9 (expected $0.997$--$1$) for  ALD MCP-PMTs and  0.84 (expected $0.8$--$0.98$) for conventional MCP-PMTs. 
The QE degradation is higher than expected; possible explanations are low-quality batches and the heat generated by electronics.
The second hypothesis was verified by measuring the quantum efficiency degradation at different environmental temperatures;
 a faster degradation was found for higher temperatures \cite{CDR}.
\begin{figure}[htp]
\centering
\includegraphics[width=4.1cm]{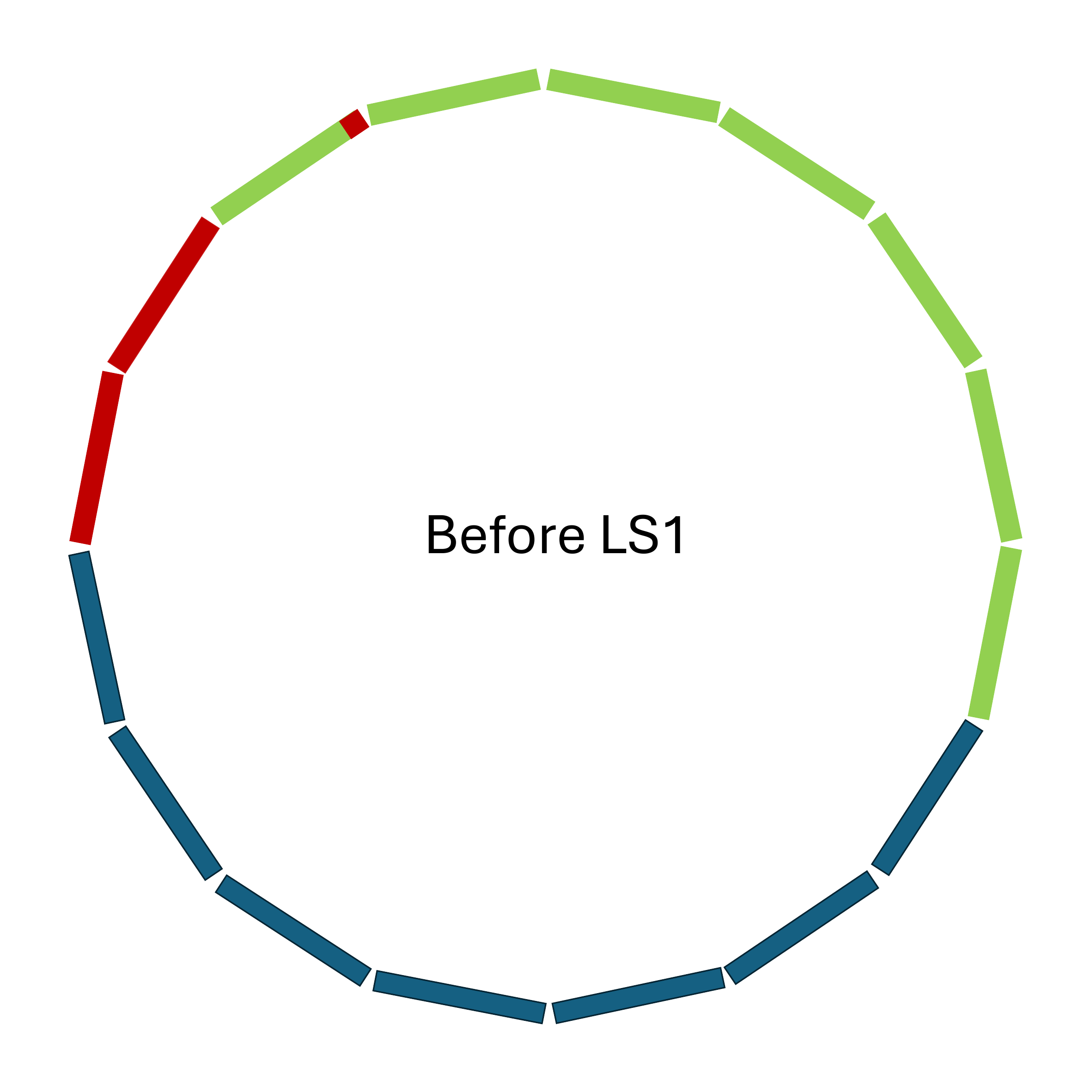}
\includegraphics[width=4.1cm]{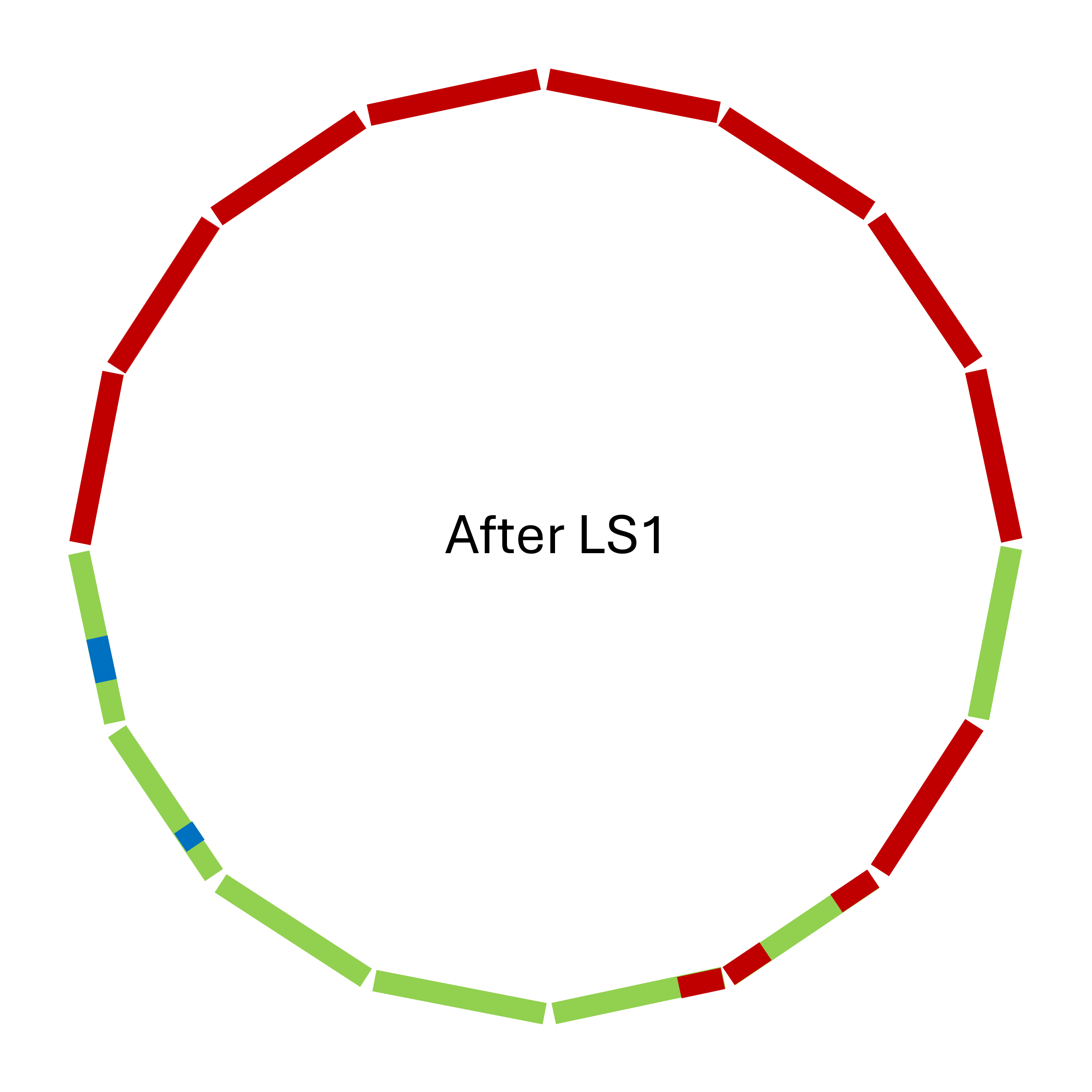}
 \includegraphics[width=4.1cm]{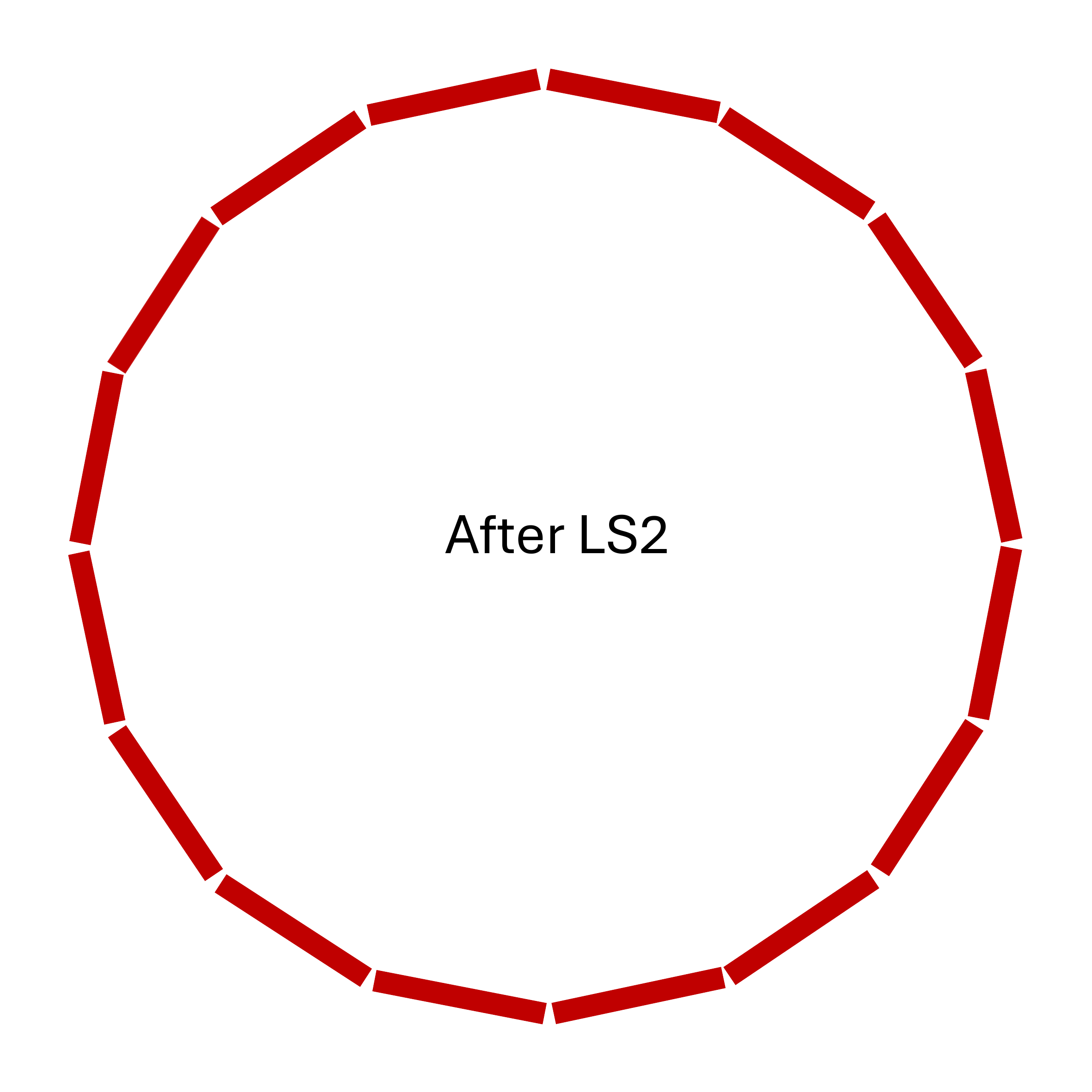}
\caption{Distribution of conventional MCP-PMTs (blue), ALD MCP-PMTs (green) and life-extended ALD MCP-PMTs (red) inside the 16 modules of the  TOP detector before LS1,
after LS1, and after the next long shutdown.}
\label{fig:f10}
\end{figure}
Figure~\ref{fig:f10} shows the distribution of different types of MCP-PMT inside the 16 modules of the TOP detector before LS1, after LS1, and the plan for the next long shutdown LS2. The production of the missing  life-extended MCP-PMTs is ongoing; 150 new MCP-PMTs have already been delivered or ordered out of 220.
The ALD MCP-PMTs have been moved from the top to the bottom slots
to have a backup option of early replacement during the annual summer break in case of faster-than-expected QE degradation.
Upgraded electronics with less power consumption and a more compact design are being considered,
as well as further lifetime improvements for MCP-PMTs or new types of photodetectors like SiPMs, in case even the life-extended ALD MCP-PMTs 
will not survive more than a year at the nominal luminosity.

\section{Conclusions}
TOP is a new concept of a compact Cherenkov detector for particle identification
that relies on multichannel long-lifetime MCP-PMTs for the precise
measurement of the arrival position and time of individual photons.
The installation of the TOP detector was completed in May 2016;
it has been successfully operating since the start of physics collisions in March 2019.
The TOP only binary PID gives $85\%$ Kaon identification efficiency with a $10\%$ pion misidentification rate.
After LS1, the fraction of active channels increased from $91$--$93\%$ to $99.5\%$.
The TOP upgrade program for the LS2 shutdown is well underway, with $68\%$ of the missing life-extended MCP-PMTs
already delivered or ordered.

\section*{Acknowledgements}
This work was supported by the following European Union's funding sources: 
Horizon 2020 Marie Sklodowska-Curie RISE project JENNIFER2 grant agreement No.~82207, 
Horizon 2020 Research and Innovation project AIDAinnova grant agreement No.~101004761.


\end{document}